\documentclass[]{aastex631}

\begin{document}

\title{Probing Ganymede's atmosphere with HST Ly$\alpha$ images in transit of Jupiter}

\author[0000-0003-0554-4691]{Lorenz Roth}
\affiliation{Space and Plasma Physics, KTH Royal Institute of Technology, Stockholm, Sweden}

\author{Gregorio Marchesini}
\affiliation{Space and Plasma Physics, KTH Royal Institute of Technology, Stockholm, Sweden}

\author{Tracy M. Becker}
\affiliation{Southwest Research Institute, San Antonio, TX, USA}
\affiliation{University of Texas at San Antonio, San Antonio, TX, USA}

\author{H. Jens Hoeijmakers}
\affiliation{Lund Observatory, Department of Astronomy and Theoretical Physics, Lund University, Lund, Sweden}

\author{Philippa M. Molyneux}
\affiliation{Southwest Research Institute, San Antonio, TX, USA}

\author{Kurt D. Retherford}
\affiliation{Southwest Research Institute, San Antonio, TX, USA}
\affiliation{University of Texas at San Antonio, San Antonio, TX, USA}

\author{Joachim Saur}
\affiliation{Universität zu Köln, Köln, Germany}

\author{Shane R. Carberry Mogan}
\affiliation{Space Sciences Laboratory, University of California, Berkeley, CA, USA}

\author{Jamey R. Szalay}
\affiliation{Department of Astrophysical Sciences, Princeton University, Princeton, NJ, USA}



\begin{abstract}
We report results from  far-ultraviolet observations by the Hubble Space Telescope of Jupiter's largest moon Ganymede transiting across the planet's dayside hemisphere. {Within} a targeted campaign on 9 September 2021 two exposures were taken during one transit passage to probe for attenuation of Jupiter's hydrogen Lyman-$\alpha$ dayglow above the moon limb. The background dayglow is slightly attenuated over an extended region around Ganymede, with stronger attenuation in the second exposure when Ganymede was near the planet's center. In the first exposure when the moon was closer to Jupiter's limb, the effects from the Ganymede corona are hardly detectable, likely because  the Jovian Lyman-$\alpha$ dayglow is spectrally broader and less intense at this viewing geometry. The obtained vertical H column densities of around $(1-2)\times 10^{12}$~cm$^{-2}$ are consistent with previous results. Constraining angular variability around Ganymede's disk, we derive an upper limit on a local H$_2$O column density of $(2-3)\times 10^{16}$~cm$^{-2}$, such as could arise from outgassing plumes in regions near the observed moon limb.
\end{abstract}



\section{Introduction}
\label{sec:intro}

Jupiter's largest moon Ganymede possesses a tenuous atmosphere produced by the solar and Jovian plasma irradiation of its icy surface \citep[e.g.][]{johnson04}. The main molecular constituents are O$_2$ \citep{hall98}, H$_2$O \citep[][]{Roth2021}, and H$_2$. For more details on Ganymede's atmosphere we refer the reader to \cite{Roth2022}.

An extended corona of atomic hydrogen was detected at Ganymede through H Lyman-$\alpha$ emissions at 1216~Å, first measured by the ultraviolet spectrometer on the Galileo spacecraft \citep{barth97}. Observations by the Space Telescope Imaging Spectrograph of the Hubble Space Telescope (HST/STIS) confirmed the H corona \citep[][]{feldman00}. The emissions are found to be globally present at a low intensity and gradually decreasing radially away from the moon, entirely consistent (only) with resonantly scattered solar Lyman-$\alpha$ radiation by an extended optically thin H corona. Analyzing HST/STIS observations taken over 15 years, \cite{alday17} found some variability in the corona brightness, yet without clear pattern such as orbital or seasonal variability. The observed radial profiles match a $1/r^2$ decrease in H number density with $r$ being the radial distance to the body center, similar to a cometary coma profile. The surface number density is about $5 \times 10^{3}$~cm$^{-3}$, the vertical (nadir) column density is about $10^{12}$~cm$^{-2}$ \citep{alday17}.   

The source for the atomic hydrogen can be dissociation of the molecular atmosphere (H$_2$O, H$_2$) or surface sputtering \citep[][]{marconi07,leblanc17}. The dissociation produces fast H atoms, populating the widely extended corona. A hydrogen corona with similar density and gradually decreasing radial profile was detected at Callisto \citep[][]{roth17-callisto}. 

On the inner-most icy moon of Jupiter, Europa, an H corona was also detected at Lyman-$\alpha$ yet with a slightly different method: The moon was observed with HST/STIS in transit of Jupiter and the H corona attenuated the Jovian Lyman-$\alpha$ dayglow background around the moon silhouette \citep{roth17-europa}. At Io, similar HST/STIS transit observations did not reveal an H corona, but showed continuum absorption by the equatorial SO$_2$ atmosphere \citep{Retherford2019}.

The Jupiter Lyman-$\alpha$ dayglow originates primarily from resonant scattering of the solar flux from the thermosphere and upwards \citep{clarke91,BenJaffel07}. The brightness of the dayglow is around 10~kR (kiloRayleigh) with a region of higher intensity near 100$^{\circ}$ Jovian longitude (called the Lyman-$\alpha$ bulge) \citep{clarke80,gladstone88,Melin2016} likely due to a broadening of the line in turbulent layers of the upper atmosphere \citep{Emerich1996}. 

Here we present observations taken with HST/STIS of Ganymede in transit of Jupiter to probe for attenuation of the Jovian Lyman-$\alpha$ background around the moon similar to the study of \cite{roth17-europa} for Europa. Our analysis focuses on the Lyman-$\alpha$ signal and we constrain Ganymede's H corona, previously measured in emission, with this complementary method. This method furthermore enables a search for absorption features directly above the limb from localized H$_2$O abundances from active outgassing sources.   

\section{Observations} \label{sec:observations}

\begin{figure}[htb]
\begin{center}
\includegraphics[width=0.7\textwidth]{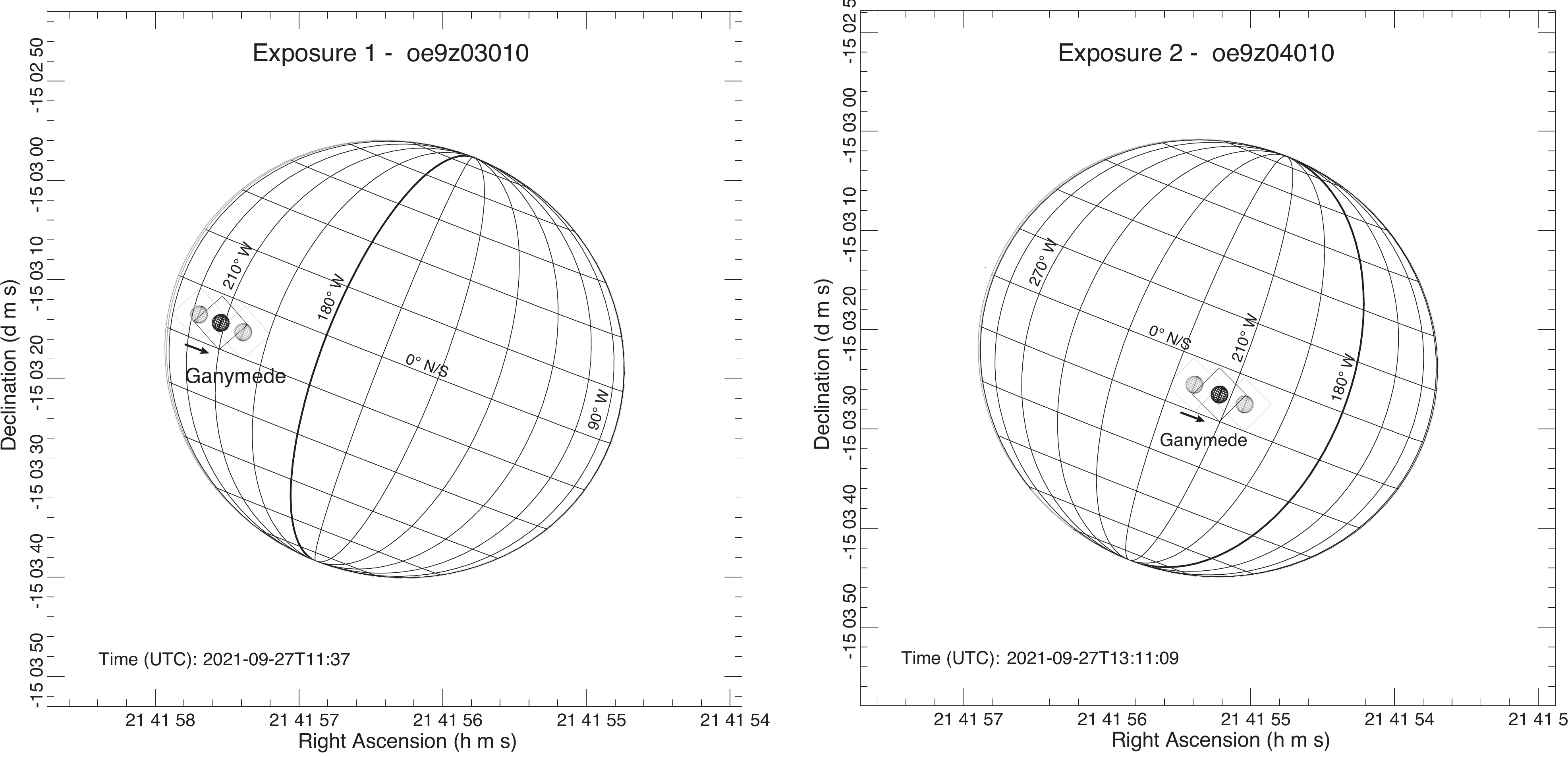}
\caption{Observing geometry of Ganymede and Jupiter during the two exposures. The Jovian longitudes and latitudes, the black moon and the black 6"x6" STIS aperture (black frame) depict the geometry at the mid point of each exposure. The positions of Ganymede and the aperture on the disk at the exposure start and end times are shown with the grey frames, with the arrow indicating the moon movement/tracking. The equator and some meridian are labeled for orientation.        
\label{fig:obsgeom}}
\end{center}
\end{figure}

On July 27, 2021 two far-ultraviolet (FUV) spectral images of Ganymede in transit of Jupiter were obtained within HST program GO 16199 using the Space Telescope Imaging Spectrograph (STIS) with the G140L grating and the rectangular 6"$\times$6" aperture box (instead of the long-slit more commonly used in previous programs). Figure \ref{fig:obsgeom} shows the aperture box, roughly centered on the moon, at the start, mid point, and end of the exposures. During the first exposure (archive root name oe9z03010) Ganymede was observed just inside the dawn limb. In the second exposure (oe9z04010) the moon was near the center of Jupiter's disk. We use the flat-fielded files for our analysis. Table \ref{tab:obsparam} summarizes the parameters of the observations. The complete transit period on that day was from 11:07 to 14:52 UTC.

\begin{table}[htb]
\caption{Parameters of the HST/STIS G140L observations of Ganymede in transit of Jupiter.}
\label{tab:obsparam}
\begin{tabular}{llcccccccc}
\hline
Visit & Date 			&   Start	& End	& 	Total 		& Cut 	    & Ganymede  & Spatial		&	Sub-observer&   Jup. SysIII		\\
 			&			&	time	& time	&	exp.time	& exp.time  & diameter	& resolution	&	W longitude & 	longitude$^a$	\\ 
			&			&	(UTC)& (UTC)	&	[min]		& [min]		& [arcsec]	& [km/pixel]	&	[$^\circ$]  & 	[$^\circ$]		\\		
\hline
oe9z03010 & 2021-07-29 & 11:16 & 11:49 &    1948.2 &   1348.2    & 1.72 & 75.1 & 179-180 & $\sim$210  \\		
oe9z04010 & 2021-07-29 & 12:43 & 13:24 &    2428.2 &   1528.2    & 1.72 & 75.1 & 182-183 & $\sim$215 \\		
\hline
\end{tabular}
\tablenotetext{a}{Refers to the System-III longitude of the projection of Ganymede on the disk of Jupiter as seen from HST (see Figure \ref{fig:obsgeom}). }
\end{table}

\begin{figure}[htb]
\begin{center}
\includegraphics[width=0.8\textwidth]{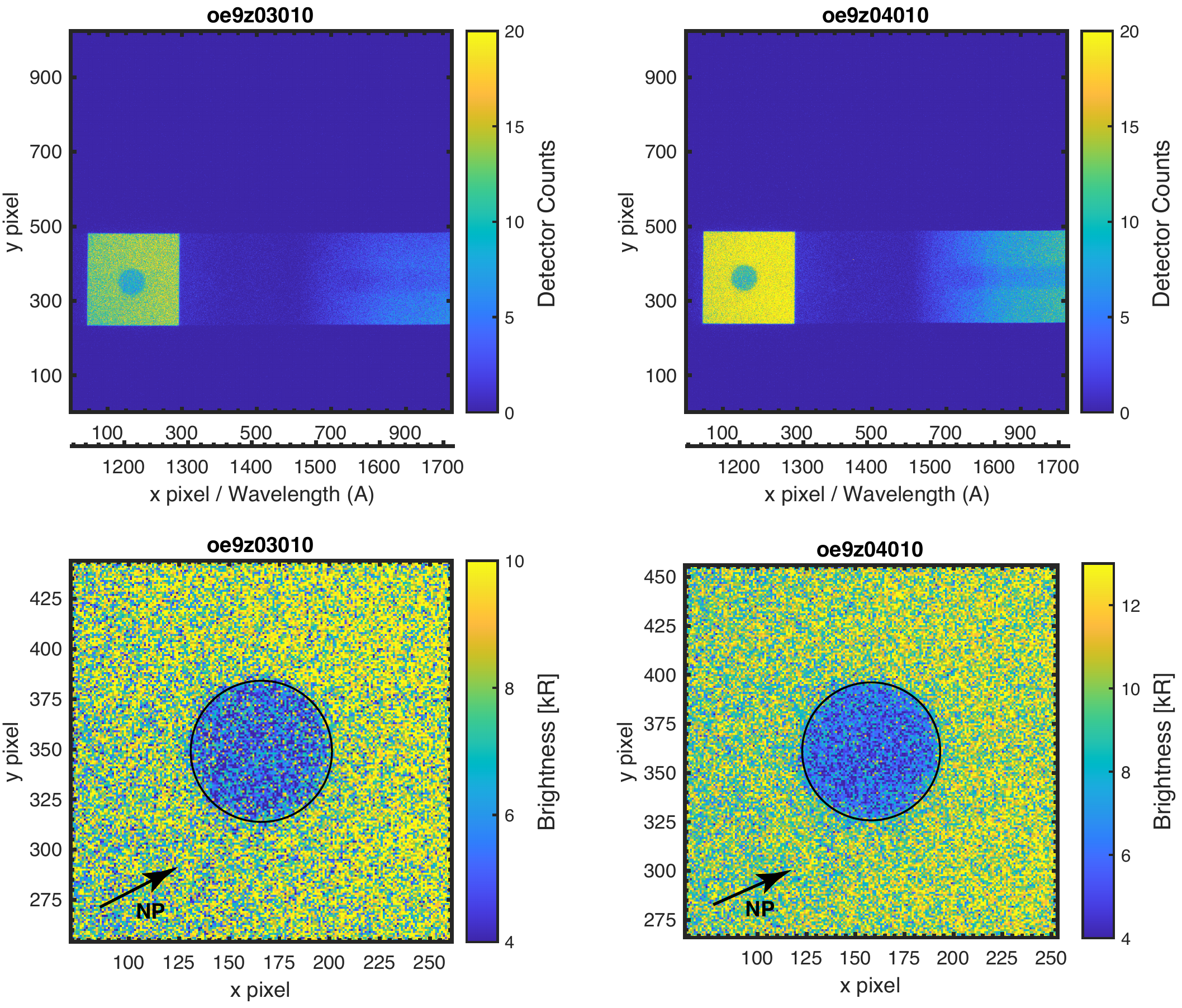}
\caption{(Top) Full STIS spectral images over the full 25"x25" detector with the dispersed signal from the 6"x6" aperture box centered near y pixel = 350. The horizontal axis contains both spatial and spectral information. Counts above and below the 6" (or $\sim$240 pixel) high trace are instrumental only. Photons in the bright box at left are interpreted to be all Lyman-$\alpha$ emissions, thus representing a monochromatic image. (Botton) Zoom-in of the Lyman-$\alpha$ image with units converted to kiloRayleigh (kR) and centered on Ganymede's disk. The direction of Ganymede's north pole (NP) is shown by the black arrow.             
\label{fig:stisimages}}
\end{center}
\end{figure}

The aperture image is dispersed by the low-resolution G140L grating and projected on the 25"$\times$25" detector, producing a 6" high spectral trace covering wavelengths between 1190 and 1720 Å (Figure \ref{fig:stisimages}, top). At wavelengths longward of $\sim$1500~Å, the Jovian H$_2$ spectrum dominated by the Lyman and Werner bands is seen with an attenuation by the transiting moon that blocks the background in the center (vertically). The aurora oxygen emissions from Ganymede at 1304~Å and 1356~Å are hardly or not discernible from the Jovian background. These weak OI~1356~Å emission surpluses indicate a similar morphology as seen in STIS images before transit \citep[][]{Marzok2022}, but are not analyzed here. The brightest box centered at 1216~A shows the H Lyman-$\alpha$ dayglow of Jupiter blocked by the disk of Ganymede.

A significant fraction of the Lyman-$\alpha$ signal on the detector originates from the Earth's geocorona. To minimize the geocorona signal we cut the first 10 and 15 minutes of the exposures, respectively, where HST is on the Earth's dayside and the geocoronal emissions are highest \citep[in analogy to the analysis in][]{roth14-apocenter}. The total and cut exposure times are given in Table \ref{tab:obsparam}.

The lower panel in Figure \ref{fig:stisimages} shows a cutout from the full detector image centered on the Lyman-$\alpha$ image of Ganymede's silhouette, with units converted to kiloRayleigh (kR, 1 kR = $4\pi\,10^9$ photons/cm$^2$/s/sr). The centering is done as part of the optimization of the forward model explained in the following section.

{The intensity measured on the disk of Ganymede is 5.3~kR and 5.6~kR in the two exposures. In order to distinguish between the contributions from foreground signal and from reflected sunlight from Ganymede's surface, we used STIS Lyman-$\alpha$ images from program 14634 (HST exposure IDs od8k40010, od8k40020) taken of Ganymede just before transit on 2 February 2017 \citep[][]{Marzok2022}. The disk-average brightness in the out-of-transit Lyman-$\alpha$ image after correction of the foreground and background was 1.2~kR. The difference in viewing geometry differs by only $\sim$6$^\circ$ longitude and can be neglected. To account for a difference in solar illumination, we then scale the value of 1.2~kR measured in 2017 with the relative intensities of the solar Lyman-$\alpha$ line for the observations ($\sim$10\% higher in July 2021) and get 1.3~kR for the estimated surface reflection brightness. This is subtracted from the brightness measured on the disk in our images.}

From the remainder of the on-disk brightness after subtracting this surface reflection, we derive the brightness of the geocorona and the interplanetary medium (IPM) hydrogen between Earth and Jupiter of 4.0~kR for exposure 1 and 4.3~kR for exposure 2. In the further analysis this signal from geocorona and interplanetary medium is subtracted from all pixels, as it can be assumed to be constant across the aperture \citep[e.g.,][]{alday17}. The approximate dayglow background at the location of the moon is given in Table \ref{tab:results} together with the other extracted brightnesses. In the following we constrain attenuation of the Jovian background around the disk by H and H$_2$O, using a simple forward model.  

\section{Analysis} \label{sec:analysis}

\begin{figure}[htb]
\begin{center}
\includegraphics[width=0.9\textwidth]{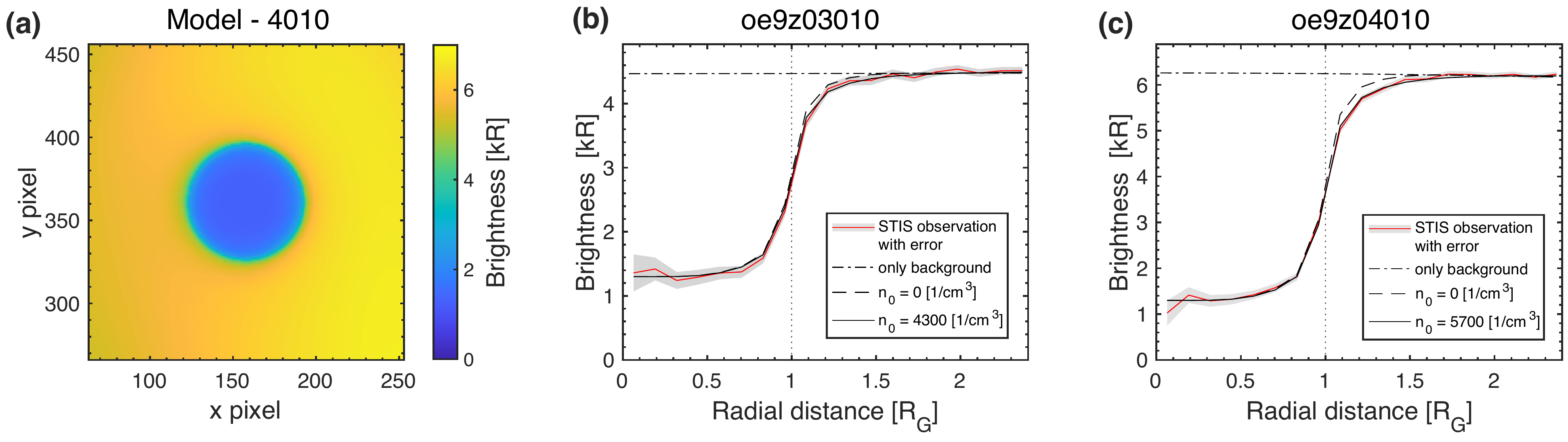}
\caption{(a) Complete model image for exposure 2 (oe9z04010) after convolution with PSF. {The 4.3~kR from geocorona and IPM are not included and the total intensity is therefore lower than in the observations shown in Figure \ref{fig:stisimages}}. (b and c) Radial brightness profiles for 3-pixel wide bins (annuli) around the disk center derived from the observation image (red) with error range (grey), from the model images without corona (dashed) and with attenuation by the best-fit corona (solid black). The fitted background (without disk of Ganymede) is shown by the dash-dotted line. Data points outside the disk (right of dotted line for disk edge) are used for fitting the corona density.       
\label{fig:model_radial}}
\end{center}
\end{figure}

For modeling the signal around Ganymede's disk, we first need to describe the unattenuated Jupiter dayglow, which reveals trends across the image due to the changes over Jupiter's disk. The trend is more pronounced in exposure 1 where the moon is close to the Jovian limb, see Figure \ref{fig:obsgeom} and bottom left panel in Figure \ref{fig:stisimages}. We fit a two-dimensional third order polynomial function to the pixels in the images that are more than 1 Ganymede radius (1 R$_G$ = 2643 km) away from the disk. We also tested a second order polynomial fit which gives similar results but in particular the x direction in the first exposure was better fit ($\sim$10\% lower chi-squared value) by a third order polynomial. The results are not sensitive to the order of the background fit. The small trend in the modelled background can be seen in the model image for exposure 2 in Figure \ref{fig:model_radial}(a). 

We then consider two effects of Ganymede's H corona on the signal:
\begin{itemize}
    \item Attenuation of the background Jovian dayglow by (isotropic) resonant scattering. Forward scattering is negligible. 
    \item Emission from resonant backscattering of the solar Lyman-$\alpha$ flux.   
\end{itemize}

The effective attenuation from resonant scattering can be described by the Beer-Lambert law as 
\begin{equation}\label{eq:beer}
I_t = I_0 \exp\left( - \sum_s \tau_s\right)   \quad,
\end{equation} 
where $I_0$ is the incident intensity (= Jovian dayglow, $I_{Jup}$) and $I_t$ the transmitted intensity. The optical depth $\tau_s$ for species $s$ is the product of the line-of-sight column density, $N_s$, and an effective cross section for the attenuation, $\sigma_s$, of the respective species, thus $\tau_s = \sigma_s N_s$.

\begin{figure}[htb]
\begin{center}
\includegraphics[width=0.8\textwidth]{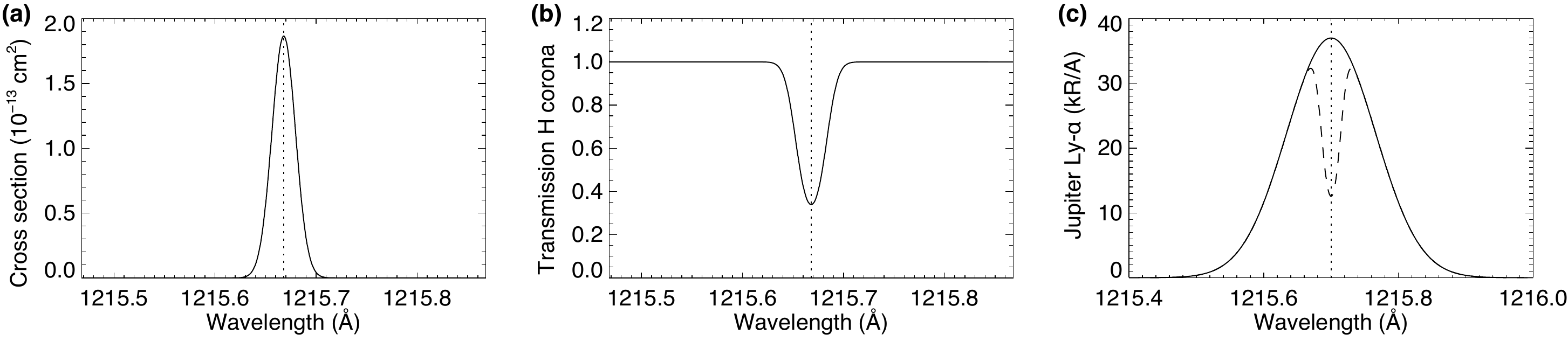}
\caption{(a) Maxwellian profile of the Lyman-$\alpha$ scattering cross section for an assumed temperature of 1000~K in Ganymede's H corona. (b) Transmission of the corona assuming a column density of $N_H = 5 \times 10^{12}$~cm$^{-2}$. (c) Assumed Maxwellian Lyman-$\alpha$ profile for the Jovian dayglow  (FWHM = 0.1~Å), before (solid) and after (dashed) attenuation by the Ganymede H corona, using the transmission in panel (b). 
\label{fig:scatter}}
\end{center}
\end{figure}

We assume that the atoms in Ganymede's corona are Maxwell distributed with a temperature of 1000~K \citep[cf. fig 5 of ][]{marconi07}. The wavelength-dependent scattering cross section \citep[e.g.,][]{chamberlain87} for H centered on the Lyman-$\alpha$ rest wavelength (1215.667 Å) for this case is shown in Figure \ref{fig:scatter}(a). The line center cross section is $\sigma_{H,0} = 1.9 \times 10^{-13}$~cm$^{-2}$. Panel (b) shows the transmission profile for the tangential (maximum) column density of $N_H = 5 \times 10^{12}$~cm$^{-2}$ \citep{alday17}. The maximum optical depth in the line center is thus $\tau = \sigma_{H,0}N_H = 0.9$ and we consider only single-scattering processes (optical thin approximation for resonant Lyman-$\alpha$ scattering).

For calculating the line-integrated cross section, the spectral characteristics of the Jovian dayglow need to be considered. The System-III longitude of Jupiter behind the moon is around 210$^{\circ}$ in both exposures (Figure \ref{fig:obsgeom}). Based on line profiles of the Jovian equatorial dayglow \citep{Bertaux80,clarke91} (away from the Jupiter Lyman-$\alpha$ bulge, which is around System-III longitude 100$^{\circ}$ W), we approximate the dayglow line shape with a Maxwellian with a full-width-half-maximum of 0.10~Å and centered at the Lyman-$\alpha$ rest wavelength (Figure \ref{fig:scatter} c). The signal after attenuation by the Ganymede H corona (transmission in panel b) is indicated by the dashed line. {By integrating the Jovian emission line profile before and after attenuation (Figure \ref{fig:scatter}c) over the line profile, we estimate an effective absorption cross section, which} for this case is $\sigma_{H,eff} = 4.9 \times 10^{-14}$~cm$^{-2}$. We use this cross section for both cases in the analysis and discuss the assumption in Section \ref{sec:discus}.

The use of a Maxwellian line profiles for the Jovian emission and Ganymede corona absorption with one temperature is a simplified description. For the case of Ganymede's H corona, the excess energies to produce H from molecular dissociation might lead to temperatures well above 1000~K \citep{huebner92}, while collisions of the fast H with molecules in the denser lowest atmosphere can reduce the temperature near the surface \citep{marconi07,Carberry2023}. 

The Jovian line profile is likely best described by a Lorentzian (for pressure broadening) or a Voigt profile (combined Doppler and pressure broadening). The Jovian emission profile is variable across the disk and with time and not known for our observations. {In particular, near the limb the profile is wider than the assumed 0.1~Å and possibly Doppler-shifted  \citep{gladstone88,Emerich1996}. We first use these approximations of the two profiles as first order estimation for the effective cross section and discuss the uncertainties and estimate the related systematic errors in Section \ref{sec:discus}.} 

The brightness from resonant backscattering of the solar flux, $I_{emis}$, can be calculated by
\begin{equation}
    I_{emis} \; [\mathrm{kR}] = 10^{-9} g_H N_H \quad,
\end{equation}
from the scattering g-factor g [photons/s] and the line-of-sight H column density $N_H$ [cm$^{-2}$] in the optically thin approximation. The integrated solar Lyman-$\alpha$ irradiance at 1 AU of $4.0\times10^{11}$~cm$^{-2}$s$^{-1}$ at the date of the observations \citep[][]{Machol2019} is related to a line center spectral irradiance of 
$3.4\times10^{11}$~cm$^{-2}$s$^{-1}$Å$^{-1}$ using the empirical formula from \cite{Kretzschmar2018}. The spectral irradiance is adjusted to the distance of Jupiter and Ganymede to the Sun of 5.0 AU at the day of the observations. The g-factor for resonant scattering of the solar Lyman-$\alpha$ flux \citep{chamberlain87} then becomes $g_H = 7.3\times10^{-5}$~s$^{-1}$. 

The radial velocity of Ganymede and Jupiter with respect to Earth is $\sim$17 km/s at the time of the observations. This corresponds to a Doppler shift of 0.7~Å at the Lyman-$\alpha$ line, well above the line widths of the Jovian dayglow, the H corona of Ganymede, and the geocorona, which are all $<$0.2~Å \citep[discussion above and][]{clarke91,alday17}. Therefore, attenuation of the emissions from Jupiter or Ganymede in the geocorona is negligible. The attenuation by the interplanetary H is also negligible with an estimated optical depth of the H column between Earth and Jupiter of $\tau < 10^{-2}$ based on our calculations after \cite{wu79}.

Because the average thermal velocity of the H at 1000~K is above the escape velocity of Ganymede (2.7~km/s), we assume an escaping H corona {with a constant radial velocity neglecting gravity and pressure. With this assumption the number density in the corona decreases with radial distance $r$ proportional to $1/r^2$ \citep{alday17}.} The line-of-sight column density outside the disk given by 
\begin{equation} \label{eq:N_H}
    N_H(r) = n_0 \pi \frac{R_G^2}{r} \quad,
\end{equation}
with the surface density $n_0$ and the radial distance to the center of the disk $r$.

The modelled brightness outside the disk of Ganymede is together given by 
\begin{equation}
    I_{off-limb} \; [\mathrm{kR}] = 10^{-9} g_H N_H(r) + I_{Jup} \exp\left( -\sigma_{H,eff} N_H(r) \right) \quad.
\end{equation}\\

For the signal on the disk ($I_{disk}$) we assume a constant brightness of 1.3 kR as derived from the out-of-transit exposures (as described in Section \ref{sec:observations}). The changing phase angle across the disk and albedo variations might affect the reflected signal \citep[e.g.,][]{ligier19} and small contributions from resonant scattering by the H corona ($\sim$0.1~kR) affect the signal but these effects are negligible and the assumed constant brightness across the disk is found to be consistent with the data. 

The effects of continuum absorption by the (sputtered or sublimated) molecular atmosphere above the terminator are expected to be negligible. The cross sections are $\sim$1.5$\times10^{-17}$~cm$^{-2}$ (OH, H$_2$O) or lower (OH, H$_2$) \citep[see figure 1 of][]{roth17-europa}. The radial column densities at the terminator are between a few times $10^{14}$~cm$^{-2}$ (O$_2$, H$_2$) and around $10^{11}$~cm$^{-2}$ (OH, H$_2$O) \citep[][]{marconi07}, for a combined optical depth of $\tau < 10^{-3}$. Even with the recently proposed denser O$_2$ atmosphere \citep{carnielli20}, the effects are negligible.  In the first step, we thus consider only the effects of the hydrogen corona. Thereafter, we constrain the presence of high local gas surpluses around the limb, as might arise at outgassing locations like for Enceladus' plumes \citep{Hansen2019} or Europa's potential plumes \citep[][]{roth14-science}. 
\\

The image is then assembled combining $I_{off-limb}$ and $I_{disk}$ and convolved with the STIS point-spread-function (PSF) \citep{krist11}.  Figure \ref{fig:model_radial}(a) shows the modelled image after the PSF convolution. \\

We then combine the pixels in radial bins (or annuli) with a radial width of 3 pixels each to quantitatively constrain the H corona, which is assumed to be radially symmetric (as defined in Equation \ref{eq:N_H}). The observed radial profiles are shown in red in Figure \ref{fig:model_radial}(b, c). A reduced average chi-square value, $\chi_\nu^2$, for the intensity in all bins outside the disk and within a radial distance of 2~$R_G$ is calculated to quantitatively compare the STIS image and model image in the region where the corona affects the data most. {In this radial bin fitting, we also varied the position of Ganymede on the detector. The fit parameters are the surface density $n_0$ and the x-y coordinate of the disk center pixel.} We retrieve the best H corona by minimizing chi-square values for each exposure, see Table \ref{tab:results}. The resulting best-fit $n_0$ for exposure 1 is about 25\% lower than the density derived for exposure 2. The obtained surface densities $n_0$ correspond to vertical (nadir) column densities, given by $N_H = n_0 R_G$ (Equation \ref{eq:N_H}), of $1.1\times10^{12}$~cm$^{-2}$ (Exposure 1) and $1.5\times10^{12}$~cm$^{-2}$ (Exposure 2).

We derive an error range around the best-fit value, setting the upper and lower limits to the surface density values where the reduced chi-squared is $\chi_\nu^2 = 1$ (Table \ref{tab:results}). For exposure 1, even a fit without corona ($n_0 = 0$) yields a chi-squared only somewhat larger than one and the error range is larger. For exposure 2, with brighter Jovian dayglow and longer exposure time (Tables \ref{tab:obsparam} and \ref{tab:results}), the stronger dependence of $\chi_\nu^2$ allows an unambiguous detection of the corona. 

Testing different line profiles for the H corona absorption and the Jovian emission and considering the hourly and daily variability of the solar flux, we estimate an uncertainty for the effective cross section and the g-factor of 20\%. The uncertainties in the two parameters mutually reinforce the change in resulting fitted densities ($n_0$). The resulting uncertainty range in density is similar to the confidence interval from the statistical errors stated in Table \ref{tab:results}; the upper limits are somewhat higher than the upper limit from confidence interval ($\sim$9$\times10^{3}$~cm$^{-3}$ compared to $\sim$8$\times10^{3}$~cm$^{-3}$).
\begin{table}[htb]
\caption{Lyman-$\alpha$ brightness of contributions and analysis results.}
\label{tab:results}
\begin{tabular}{llcccccccc}
\hline
Exposure 	&	 Surface & Geocorona	& Jupiter &	 Fitted		    & $\chi_\nu^2$	& $\chi_\nu^2$  &  & {Upper limit H$_2$O }		\\
 			& reflection & \& IPM  &  dayglow&  $n_{H,0}$ $^a$	& best-fit	& no       &  & {plume column density}		\\ 
	        &	(kR)	     & (kR)		&	(kR)   &(cm$^{-3}$)     & corona	& corona   &  & {(cm$^{-2}$)} 	\\		
\hline
oe9z03010 &		1.3		&	4.0	&   4.6  & $4.3^{+3.5}_{-3.1}\times10^{3}$	& 0.8  & 1.2 & & $3.0 \times10^{16} $	\\		
oe9z04010 &		1.3		&	4.3	&	6.3 &  $5.7^{+2.6}_{-2.4}\times10^{3}$	& 0.7  & 2.5 & & $2.1 \times10^{16} $	\\
\hline
\end{tabular}
\tablenotetext{a}{Used all pixels outside the disk for fit. The range reflects a confidence interval where $\chi_\nu^2 < 1$.}
\end{table}

\begin{figure}[ht!]
\begin{center}
\includegraphics[width=0.5\textwidth]{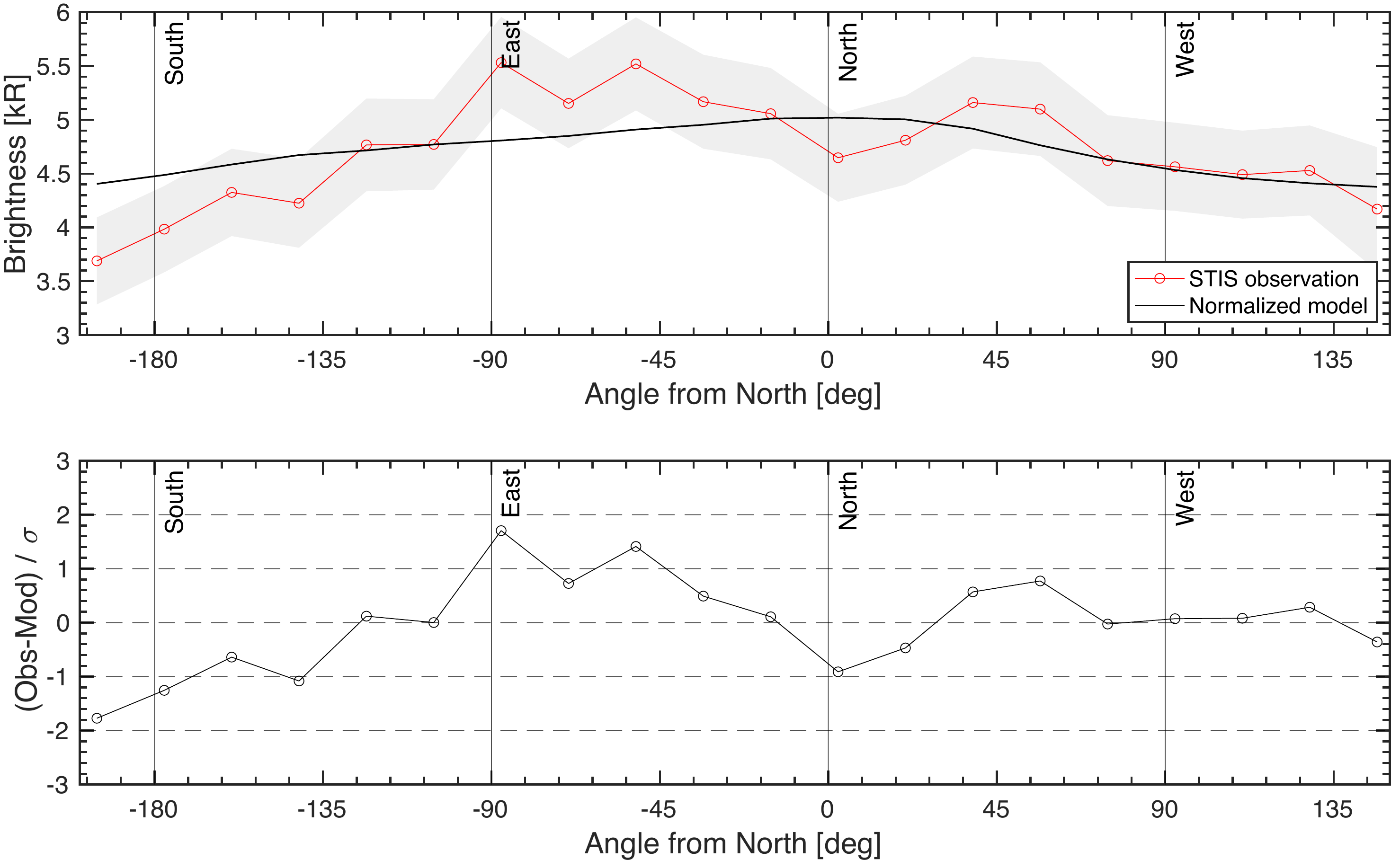}
\caption{(Top) Brightness of the 24 limb bins around Ganymede for observation oe9z04010 (red) with statistical error (shaded grey) and for our model image as shown  in Figure \ref{fig:model_radial}a (black line). (Bottom) Difference between model and observation divided by the error $\sigma$. 
\label{fig:limbplume}}
\end{center}
\end{figure}

To search for local limb anomalies, we divide the limb area between 1.0 and 1.1~$R_G$ (altitude 0--260~km) in 20 bins with an angular width of 18$^{\circ}$ each. Each bin covers an area of about $220\times10^3$~km$^2$ at Ganymede. The brightness of the model image is then normalized such that the mean model brightness of all pixels between 1.0 and 1.2~$R_G$ matches that of the observation. Then we calculate the average brightness in each bin in the observation image and the model image. Figure \ref{fig:limbplume} shows the comparison for exposure 2, which has higher SNR as discussed above. The trend around the disk originates primarily from the trend in the dayglow background (see Figure \ref{fig:stisimages} and \ref{fig:model_radial} right) with additional small effects from the anisotropic PSF. The lower panel shows the statistical significance of the deviation of the measured bin brightness from the model. The spreading of these deviations represents normally distributed statistical fluctuations and possibly shows that the background variation is not fully reflected in our model fit. No significant outliers are thus present. The variation around the limb for exposure 1 is similarly consistent with only statistical fluctuations and the largest outlier is $<$2$\sigma$.

A localized gas abundance in one bin would decrease intensity, described by Equation \ref{eq:beer}. To set an upper limit on water vapor plumes, we define a 3$\sigma$ outlier limit and calculate the maximum attenuation of the background $I_0$ in this limb region as $I/I_0 = (I_0 - 3\sigma)/I_0$. With the H$_2$O absorption cross section of $\sigma_{H_2O} = 1.5 \times 10^{-17}$~cm$^{2}$, we calculate an upper limit for the column density as
\begin{equation}
    N_{H_2O}^{max} = \frac{-\ln((I_0 - 3\sigma)/I_0)}{ \sigma_{H_2O} } \quad,
\end{equation}
for each exposure, listed in Table \ref{tab:results}. Limits on localized abundances for other possible molecular species are even higher ($N_{H_2} \sim 10^{19}$~cm$^{-2}$, $N_{O_2} \sim 10^{20}$~cm$^{-2}$), because the respective cross sections are lower at Lyman-$\alpha$ \citep{roth17-europa}. 

\section{Discussion}\label{sec:discus}

The brightness of the Jupiter dayglow background around Ganymede is somewhat higher in the second exposure (6.3~kR) than in the first (4.6~kR), consistent with the higher dayglow found in the disk center compared to the limb \citep[][]{clarke91}. Overall, the dayglow brightness is lower than previous measurements \citep{Bertaux80,clarke91,Melin2016} but similar to the recent results from \cite{roth17-europa}.  The comparably low dayglow brightness might be partially explained by the rather low solar intensity, but that cannot fully account for the difference of a factor 2 or more. 

The obtained H surface densities (Table \ref{tab:results} and corresponding column densities of $1.1\times10^{12}$~cm$^{-2}$ (Exp. 1) and $1.5\times10^{12}$~cm$^{-2}$ (Exp. 2) are consistent with the HST/STIS observations of the H corona emissions (out of transit) \citep[][]{feldman00,alday17}. Considering both systematic (Table \ref{tab:results}) and statistical errors of the values derived here, the HST/STIS corona constraints from emissions observed by HST \citep{feldman00,alday17} are more reliable and precise than those obtained here. Hence, our results confirm the earlier detected corona through a complementary method but do not provide additional quantitative constraints.  
The HST densities are $\sim$6$\times$ lower than the early constraints from Galileo UVS data \citep[][]{barth97}, which are however less well calibrated. {If not instrumental, this difference could possibly relate to the different viewing geometries: HST always observes the corona near the terminator at small solar phase angles, while Galileo measured the H corona closer to the sub-solar point at a solar phase angle of 60$^\circ$. If, for example, sublimated H$_2$O is an additional source for H on the dayside \citep{Carberry2023}, then the H abundance might indeed be higher in the region closer to the sub-solar point probed by Galileo. A global change in the H corona by a factor of 6 (as explanation for the difference between the Galileo and HST results) is not seen between the different HST data taken from 1998 to 2014 of the H corona \citep{alday17} nor of the oxygen atmosphere \citep{Marzok2022} and is thus unlikely.}

The weaker attenuation and $\sim$25\% lower derived H density in the first exposure, where Ganymede is observed near the limb (Figure \ref{fig:obsgeom}), {is likely} due to the broader and possibly Doppler-shifted Lyman-$\alpha$ line of the Jovian dayglow in this region \citep[cf. figure 2 of][]{clarke91}. A broadening of 25\% or a shifted of the line center by 0.04~Å (9 km/s) would lead to a $\sim$25\% lower effective absorption by Ganymede's H corona (keeping the corona temperature and density unchanged). We note, however, that the difference in derived H densities between the exposures is much smaller than the error.  

Although we assumed a radially symmetric H corona, the observations mostly probe the regions above the observed limb, i.e., above the terminator of Ganymede for the observing geometry from Earth. Our obtained vertical column density of (1.1-1.5)$\times10^{12}$~cm$^{-2}$ is 4-5 times higher than the vertical column density found in the atmosphere simulations of \cite{marconi07} near the terminator of 0.3$\times10^{12}$~cm$^{-2}$ (their figure 3, at 90$^\circ$ sub-solar longitude). Other simulations \citep{turc14,leblanc17} find even lower H densities, although \cite{turc14} explains this partly by reactions not considered in their model. At the sub-solar point where dissociation of sublimated H$_2$O is an additional source for H in the simulation of \cite{marconi07}, their simulated column density is similar to our values. This can, however, not account for the higher H column density above the limb measured here. Recently, \cite{Carberry2023} showed that dissociation of a global H$_2$ atmosphere with surface densities of $n_{H_2} \sim (4-10)\times 10^{7}$~cm$^{-3}$, which is 1-2 orders of magnitude higher than the H$_2$ density in the simulations of \cite{marconi07} and \cite{leblanc17}, can sustain the H corona at Callisto which has similar density and distribution. 

Loss of neutrals from Ganymede potentially also creates a cloud or torus in the moon's orbit \citep[][]{Huang1987}. \cite{marconi07} estimates escape rates of $3.0\times 10^{26}$~H/s and of $1.3\times 10^{27}$~H$_2$/s, which might yet be higher if both H and H$_2$ atmospheres from \cite{marconi07} indeed need to be up-scaled in density to agree with the observational constraints on the H corona (see discussion in previous paragraph). On the other hand, observations of H$_2^+$ pick-up ions by the Juno spacecraft did not reveal a substantial source of the ions at Ganymede's orbit limiting the H$_2$ neutral density and thus loss rate to $2.0\times 10^{26}$~H$_2$/s (50\% of Europa's rate) or lower \citep{Szalay2022}. Taken together, the H corona density suggests H$_2$ loss rates from the atmosphere (via simulations) that are more than one order of magnitude higher than the constraints from H$_2^+$ pick-up ions, suggesting that the pathways of gaseous neutral hydrogen species from the icy moons are not fully understood yet.

Our obtained H vertical column densities for Ganymede are about five times higher than the H column densities derived for Europa with the same method \citep[][]{roth17-europa}. However, in \cite{roth17-europa}, the H abundances were calculated using the line center scattering cross section for H instead of an effective line-integrated cross section. The difference between the line center cross section used in \cite{roth17-europa} and the cross section used here is also about a factor of five. Because the line-integrated cross section is the more appropriate value and because our results here are consistent with H densities measured in emission for Ganymede \citep{alday17}, our analysis suggests that Europa's H corona is about five times denser than derived in  \cite{roth17-europa}. Constraints on Europa's corona through Lyman-$\alpha$ emission observations can help to resolve this. If the H density at Europa is higher, this would also suggest a conflict between H$_2$ escape rate constraints based on H corona densities and those based on H$_2^+$ ions measurements \citep{Szalay2022}. 

We provide the first upper limits for for localized (plume) H$_2$O column densities at Ganymede of $(2-3) \times10^{16}$~cm$^{-2}$. Given the transit geometry the limits apply only to plumes located at or near 90$^{\circ}$ and 270$^{\circ}$ W longitude (near the observed limb). The obtained H$_2$O upper limits on the column density are about twice the column densities derived for Enceladus' plumes in occultation \citep[][]{hansen2020} and similar to those derived for plumes at Europa from H$_2$O auroral emissions \citep[][]{roth14-science}. Our limits are an order of magnitude lower than those for Europa derived from putative absorption features in STIS FUV filter (around 1600 Å) images \citep[][]{sparks16}, but these features are also consistent with statistical fluctuations only \citep[][]{giono20}. It was shown for Enceladus that the detection of H$_2$O plumes through attenuation of the planet's dayglow background is more sensitive at Lyman-$\alpha$ compared to longer FUV wavelengths \citep{Hansen2019}.

Our column density limits are derived for the area of one limb bin ($220\times10^3$~km$^2$). The product of the column density limit and bin area gives an upper limit on the total number of molecules of $\sim5\times10^{31}$ or total mass of $1.5\times10^{6}$~kg, similar to the amount of H$_2$O for an outgassing event at Europa derived from large aperture infrared emissions \citep{paganini19}. 

Our results are likely the best constraints on localized H$_2$O on Ganymede available for now. {Results from the first James Webb Space Telescope observations of Ganymede taken earlier in 2022 might provide new insights soon.} The Ultraviolet Spectrograph on the JUpiter Icy Moon Explorer (JUICE) mission{, almost ready for its launch in 2023,} aims to provide global constraints on plumes through stellar and solar occultation and Jupiter transit measurements, during the cruise and Ganymede orbit phases \citep{Grasset2013}.  

\section{Summary}

Two spectral images of Ganymede in transit of Jupiter were taken with HST/STIS on July 29, 2021. In this paper, we analyze the Lyman-$\alpha$ signal in the images constraining the effects of Ganymede's tenuous atmosphere on the background Jovian dayglow above the limb of the moon disk. To constrain atmospheric attenuation, we construct a simple forward model that accounts for effects from the H corona and from localized high H$_2$O abundances. Fitting the model to the data we get the following results:
\begin{itemize}
    \item The presence of an extended H corona around Ganymede is confirmed.
    \item Fitted H densities are similar to previous results, however with larger uncertainties.
    \item No signs of localized anomalies are found and an upper limit for H$_2$O plume abundance is derived.
\end{itemize}

Finally, our analysis suggests that the cross section for the study of Europa in transit by \cite{roth17-europa} was overestimated by a factor of five and the H densities at Europa therefore underestimated. If Europa's H corona is indeed denser, then all three icy moons of Jupiter have H coronae with similar column densities of a few times $10^{12}$~cm$^{-2}$ and similarly gradual density profiles. The similarities suggest similar generation (e.g., dissociation of hydrogen-bearing molecules in the atmosphere) and loss processes (e.g., return to surface and escape from atmosphere). As discussed earlier, \cite{Carberry2023} showed that a global H$_2$ atmosphere could produce the observed H corona at Callisto. Therefore, the similar densities and distributions observed at Europa and Ganymede might also be indicative of an H$_2$-source, which is indeed plausible given that all three bodies have icy surfaces in which radiolysis should produce H$_2$. Moreover, such a suggestion is also consistent with the recent Juno flyby that detected H$_3^+$ \citep{Allegrini2023}, which is produced via chemistry in an H$_2$ atmosphere.

Even with similar loss rates at the three moons, a neutral torus would however be densest at Europa, because the sources are dispersed in larger volumes in the orbits of the outer moons Ganymede and Callisto. Further studies of the atmospheres and losses are needed to understand the characteristics of the atomic H coronae and their relations to the molecular atmospheres and possible neutral tori or clouds at Ganymede and the other moons.\\

\section*{Acknowledgments}
The authors thank John Clarke for discussions and input on the line profile of the Jovian dayglow. L.R. appreciates support from the Swedish National Space Agency through grant 2021-00153 and from the Swedish Research Council through grant 2017-04897. S.R.C.M. was supported by Solar System Workings (SSW) grants 80NSSC21K0152 and NNX16AR99G.

All of the data presented in this paper were obtained from the Mikulski Archive for Space Telescopes (MAST) at the Space Telescope Science Institute. The specific observations analyzed can be accessed via \dataset[10.17909/de19-9z85]{https://doi.org/10.17909/de19-9z85}.

%

\vspace{5mm}
\facilities{HST/STIS}




\bibliography{bib2022}{}
\bibliographystyle{aasjournal}



\end{document}